\documentclass[aps,prl,amsmath,amssymb,amsfonts,twocolumn,notitlepage,nofootinbib]{revtex4-1}
\usepackage{enumerate}
\usepackage{graphicx}
\usepackage{dcolumn}
\usepackage{bm,bbm}
\usepackage{amsmath}
\usepackage{amssymb}
\usepackage{color,hyperref}

\usepackage{soul}
\usepackage[utf8]{inputenc}

\newcommand{\sz}[1]{\sigma^z_{#1}}
\newcommand{\spl}[1]{\sigma^{+}_{#1}}
\newcommand{\smi}[1]{\sigma^{-}_{#1}}

\newcommand{\ket}[1]{|{#1}\rangle}
\newcommand{\bra}[1]{\langle {#1}|}
\newcommand{\ta}{\tilde{a}}
\newcommand{\td}[1]{\tilde{#1}}
\newcommand{\Li}{\mathcal{L}}
\newcommand{\pp}{\partial}
\newcommand{\D}{\mathcal{D}}

\newcommand{\dg}{\dagger}

\newcommand{\tr}{\mathrm{tr}}

\newcommand{\arccosh}{\mathrm{arccosh}}
\newcommand{\av}[1]{\langle{#1}\rangle}

\newcommand{\minus}{-}

\newcommand{\be}{\begin{equation}}
\newcommand{\ee}{\end{equation}}

\newcommand{\bea}{\begin{eqnarray}}
\newcommand{\eea}{\end{eqnarray}}

\def\Pm{\mathbb{P}}
\def\Tm{\mathbb{T}}

\def\ave#1{\langle #1 \rangle}

\def\tit#1{{\em #1},}

\newcommand*{\affaddr}[1]{#1} 
\newcommand*{\affmark}[1][*]{\textsuperscript{#1}}

\begin{document}

\title{Exact Bethe ansatz spectrum of a tight-binding chain with dephasing noise}

\author{%
Mariya V. Medvedyeva\affmark[1], Fabian H. L. Essler\affmark[2], and Toma\v z Prosen\affmark[1]\\
\affaddr{\affmark[1] {\it Physics Department, Faculty of Mathematics and Physics, University of Ljubljana, Ljubljana, Slovenia}}\\
\affaddr{\affmark[2] {\it The Rudolf Peierls Centre for Theoretical Physics, Oxford University, Oxford, OX1 3NP, United Kingdom}}
}

\begin{abstract}
We construct an exact map between a tight-binding model on any
bipartite lattice in presence of dephasing noise and a Hubbard 
model with imaginary interaction strength. In one dimension, the exact
many-body Liouvillian spectrum can be obtained by application of
the Bethe ansatz method. We find that both the non-equilibrium steady state
and the leading decay modes describing the relaxation at late times
are related to the $\eta$-pairing symmetry of the Hubbard model. We
show that there is a remarkable relation between the time-evolution of an
arbitrary $k-$point correlation function in the dissipative system and 
$k$-particle states of the corresponding Hubbard model. 
\end{abstract}

\maketitle

{\it Introduction.--} The coupling to the environment often has a
non-negligible influence on a many-particle system, and may drive it
to a \emph{non-equilibrium steady state} (NESS), that is different
from the ground or thermal equilibrium states. Within the so-called
Markovian description, assuming that the internal bath dynamics is
much faster than that of the system so there is no back action of the
system onto its environment, one has a well defined mathematical
description of open many-body systems in both classical and
quantum contexts. In the quantum realm, the open system's Liouvillian
dynamics is described by 
the Lindblad master equation \cite{Breuer} for the time-dependent
density matrix. A standard way of analyzing the Lindblad equation
is by means of perturbative methods~\cite{perturbation,DiehlKeldysh},
but it is highly desirable to have exact solutions in specific representative cases.
While NESSs have been constructed exactly in a number
of cases, in both classical \cite{ASEPNESS} and quantum
settings \cite{Pros2008,Marko,ProsenExact}, solving the full dynamics,
i.e. diagonalizing the Liouvillian, for any nontrivial many-body
system is a formidable task. In the quantum case this has so far been
possible only for noninteracting systems \cite{Pros2008}. On the other
hand, in certain classical stochastic many body systems like the
asymmetric simple exclusion processes, the full Markov chain
can be diagonalized in terms of the Bethe ansatz.
\cite{ASEPBA} 
It is then natural to ask whether there are quantum many-body dissipative systems
that are Bethe ansatz solvable.

In this Letter we present an exactly solvable dissipative many-body
quantum system that is not equivalent to a free theory: a fermionic
tight-binding model on a bipartite lattice with dephasing noise. In
one spatial dimension this model is equivalent, up to boundary
conditions, to a dephased spin-1/2 $XX$ chain. The model has
applications to ultra-cold atoms in an optical lattice subjected
to light scattering~\cite{Sarkar14}, and to superconducting flux 
qubits coupled to a fluctuating electromagnetic
environment~\cite{qbits}. The dissipation in the form of dephasing 
destroys the quantum coherence, i.e. off-diagonal density matrix elements
in the Fock space basis. It is known that such models exhibit
diffusive behavior~\cite{Esp2005,Eis2011}.  

Even though the Hamiltonian of our model is quadratic in fermion
operators, the dissipative term leads to quartic terms in the
effective evolution operator,  which renders its diagonalization
a non-trivial task. We put 
forward a simple unitary transformation, within the thermofield
description \cite{thermo}, that maps the Liouvillian superoperator of
our model on an arbitrary bipartite lattice to the Hubbard Hamiltonian
with imaginary interaction strength. In one spatial dimension this
means that the entire machinery of the Bethe ansatz formalism
\cite{bookHub} is applicable: we can obtain the full Liouvillian
spectrum by solving the Bethe-ansatz equations, as well as reconstruct
the time-evolution of the density matrix by expanding it over the
Bethe-ansatz wave-functions.  

{\em Dephasing model.--}
We consider dissipative many-body dynamics of free fermions on a bipartite
lattice in the tight-binding approximation with Hamiltonian 
\be  
H=\sum_{\langle j,k\rangle} (a_j^\dg a_k +a^\dg_{k} a_{j}).\label{Htb}
\ee
Here $a_j^\dg/a_j$ are fermionic creation/annihilation operators on
site $j$, and $\langle j,k\rangle$ denote nearest-neighbour links
connecting the two sublattices $A$ and $B$. The
dissipative dynamics is described by the Lindblad equation $\frac{\pp
  \rho}{\pp t} = \Li[\rho]$, where 
\begin{equation} 
\Li [\rho] = -i [H,\rho]+\D[\rho],~ \D[\rho] = \sum_j \left(2l_j \rho l_j^\dagger  - \{ l_j^\dagger l_j,\rho\} \right).
\label{eq:Lin}
\end{equation}
The Lindblad operators are $l_j=\sqrt{2\gamma}a^\dg_j
a_j$. The derivation of this dissipation term follows the books~\cite{Gardiner,Breuer}, or Ref.~\cite{Sarkar14} in the optical
lattice context. The dephasing strength $\gamma$ is a function of the
laser intensity and detuning.  

It is useful to express the generator of the time-evolution (Liouvillian) in the
thermofield representation \cite{thermo}. To that end we introduce a second
set of fermionic operators $\ta_j^\dg$ and $\ta_j$, which act on the
density matrix by right
multiplication~\cite{Dzh2012,fermionicHS}. The Liouvillian then takes
the form
\be 
\Li = -i \mathcal{H} + 2\gamma \sum_j \left(2 a^\dg_j
a_j \ta^\dg_j \ta_j - a^\dg_j a_j - \ta^\dg_j
\ta_j\right)\label{LiTB},
\ee
where $\mathcal{H}=H - \tilde{H}$ is the time-evolution generator
of the closed system in the thermofield representation.  
We note that the total numbers of particles of each `flavour'
(non-tilde and tilde operators) $M_1 = \sum_j a^\dg_j a_j$,  $M_2 =
\sum_j \ta^\dg_j \ta_j$ are conserved during
time-evolution, as is their sum $N=M_1+M_2$. In the following we will
choose $M_1\equiv M$ and $N$ as good quantum numbers of our problem.

{\em Transformation to imaginary Hubbard model.--} We now perform a
unitary transformation which flips the sign of the tilde-Hamiltonian 
\be{\mathcal U}  = \prod_{j\in { A}} e^{i \pi \ta_{j}^\dg \ta_{j}}=
\prod_{j\in { A}} (1-2 \ta_{j}^\dg \ta_{j}).
\label{transformTB}
\ee
In the transformed basis the generator of time-evolution can be
written as the Hamiltonian of the Hubbard model at a finite chemical
potential and imaginary interaction strength $u = i \gamma$
\begin{eqnarray}
H_{\mathrm{Hubb}}&\equiv& i {\mathcal U}^{\dg}\Li {\mathcal U}^{} 
= \sum_{\langle j,k\rangle;\sigma = \uparrow,\downarrow} \left(c^\dg_{j,\sigma}
c_{k,\sigma} +h.c.\right) +\nonumber\\ &+& 4 u \sum_j
n_{j,\uparrow}n_{j,\downarrow}
- 2 u \sum_{j;\sigma = \uparrow,\downarrow}n_{j,\sigma}.
\label{ImHubb}
\end{eqnarray} 
Here the fermionic operators for the spin-up and spin-down are related
to the normal and tilde operators of the dephasing model by
$c_{i,\uparrow} = a_i$, $c_{i,\downarrow} = \td{a}_i$, and
$n_{j,\sigma}=c^\dagger_{j,\sigma}c_{j,\sigma}$. The imaginary-$u$
Hubbard model (\ref{ImHubb}) exhibits an $SO(4)$
symmetry \cite{HL,etaPair,bookHub}. The generators of the 
constituent $\eta$-pairing $SU(2)$ algebra are
$\eta^z =\sum_{j} (n_{j,\uparrow}+n_{j,\downarrow}-1)$, $\eta^+
=\sum_{j\in A} c^\dagger_{j,\downarrow}c^\dagger_{j,\uparrow}-
\sum_{j\in B} c^\dagger_{j,\downarrow}c^\dagger_{j,\uparrow}=
(\eta^{\minus})^\dagger$, and will play an important role in the following.

{\it Steady State.--} The NESS is characterized by the condition $\Li
\rho = 0$. In the sector $L=N=2M$, where $L$ is the total number of sites, 
it is easy to read off the NESS in the Hubbard model
representation 
\be
|{\rm NESS}\rangle=\big(\eta^\dagger\big)^M|0\rangle,
\label{NESS}
\ee
where $|0\rangle$ is the fermion vacuum defined by
$c_{j,\sigma}|0\rangle=0$. Given that $H_{\rm Hubb}|0\rangle=0$, the
$\eta$-pairing symmetry implies that the state (\ref{NESS}) has zero
eigenvalue as well and therefore is a steady state. 
$\eta$-pairing states like (\ref{NESS}) attracted attention in
the early nineties in relation to high-$T_c$ superconductivity,
because they are exact eigenstates of the Hubbard Hamiltonian that
display off-diagonal long-range order\cite{etaPair}. However, in the
Hubbard model they can never be ground states. The corresponding
state in the dissipative model is obtained by undoing the unitary
transformation (\ref{transformTB}), and is of the form $\sum_{n_j\in\{0,1\}}
\ket{n_1,\ldots,n_L}_{\uparrow} \otimes
\ket{n_1,\ldots,n_L}_{\downarrow}$, where $\ket{n_1,\ldots,n_L}$ run
over all Fock states of one flavour. Hence the density matrix
corresponding to the NESS is an identity operator and represents a
completely mixed (infinite temperature) state.  

{\it Correlation function -- wave-function duality.--} The evolution
of the expectation value of the operator $O$ obeys the
equation 
\be \frac{d\av{O}}{dt}= \tr \left(\frac{d\rho}{dt} O \right) = \tr (\Li [\rho] O). \label{evolutionO}\ee
A straightforward calculation (substituting the explicit form of
$\Li [\rho]$ in Eq.~(\ref{evolutionO}) and using cyclic permutation invariance under the trace) shows that the equation for the dissipative
time-evolution of the $2k$-point correlation function  
$$G^{m_1\ldots m_k}_{n_1\ldots n_k}(t) =\tr(a^\dg_{m_1}\ldots a^\dg_{m_k} a_{n_1}\ldots a_{n_k}\rho(t))$$ 
is given by the same equation as the evolution of the density matrix
elements corresponding wave functions in the $2k$ particle sector
$\Psi^{m_1\ldots m_k}_{n_1\ldots n_k}\equiv G^{m_1\ldots m_k}_{n_1\ldots n_k}$,
$$\rho(t) = \!\!\!\!\!\!\sum_{m_1\ldots m_k n_1 \ldots n_k}\!\!\!\! \Psi^{m_1\ldots m_k}_{n_1\ldots n_k}(t) a^\dg_{m_1}\ldots a^\dg_{m_k} \ket{0}\bra{0} a_{n_1}\ldots a_{n_k}.$$
This duality, combined with the integrability of the imaginary-$u$
Hubbard model, gives a simple way for calculating general correlation
functions of the tight-binding model with dephasing.

It has been noted previously~\cite{Eis2011} that a one dimensional
tight-binding model with (different) dephasing gives rise to a closed
system of equations for correlation functions up to a given order, but
in our case the Bethe ansatz solvability makes the duality much more
powerful. 



{\it Liouvillian spectrum in one dimension.--}
The fact that the interaction strength is purely imaginary does not
spoil the algebraic integrability structure of the Hubbard model. 
In fact, the Bethe ansatz wave-functions (related to the system's
density matrix) and the Bethe ansatz equations (BAE) are simply
obtained from the regular Hubbard model by taking the interaction
strength to be imaginary. The BAE for our case read~\cite{BetheHub}:
\begin{eqnarray}
&&e^{i k_j L} = F_1 \prod_{\alpha=1}^{M} \frac{\Lambda_\alpha-\sin
    k_j+\gamma}{\Lambda_\alpha-\sin k_j-\gamma},~j=1,\ldots,N, \nonumber\\
&&\prod_{j=1}^{N} \frac{\Lambda_\alpha-\sin k_j+\gamma}{\Lambda_\alpha-\sin k_j-\gamma} = F_2 \prod_{\beta,\beta\ne \alpha} \frac{\Lambda_\beta - \Lambda_\alpha + 2\gamma}{\Lambda_\beta - \Lambda_\alpha - 2\gamma},~ \label{BAE}\\
&&\alpha=1,\ldots,M.\nonumber
\end{eqnarray}
Here $k_j$ and $\Lambda_\alpha$ are rapidities corresponding to
charge and spin excitations, and we have introduced phase
factors $F_{1,2}$ for later convenience. In the case at hand we have
$F_1=F_2=1$. The eigenvalues of the Liouvillian ${\cal L}
\rho_\epsilon = \epsilon \rho_\epsilon$ corresponding to a given
solution of the Bethe equation $\{k_j\}$ are
\be 
\epsilon (\{k_j\}) =-2i\sum_{j=1}^N \cos k_j - 2 N\gamma. 
\label{decays}
\ee
While the BAE allow us in principle to determine the full spectrum of
the Liouvillian, we will be mainly interested in the structure of the
slowest decaying NESS-excitations \cite{NESSExcitation}, i.e. eigenvalues with
the largest real parts. 

\begin{figure}[ht]
\begin{center}
\includegraphics[width=0.95\linewidth]{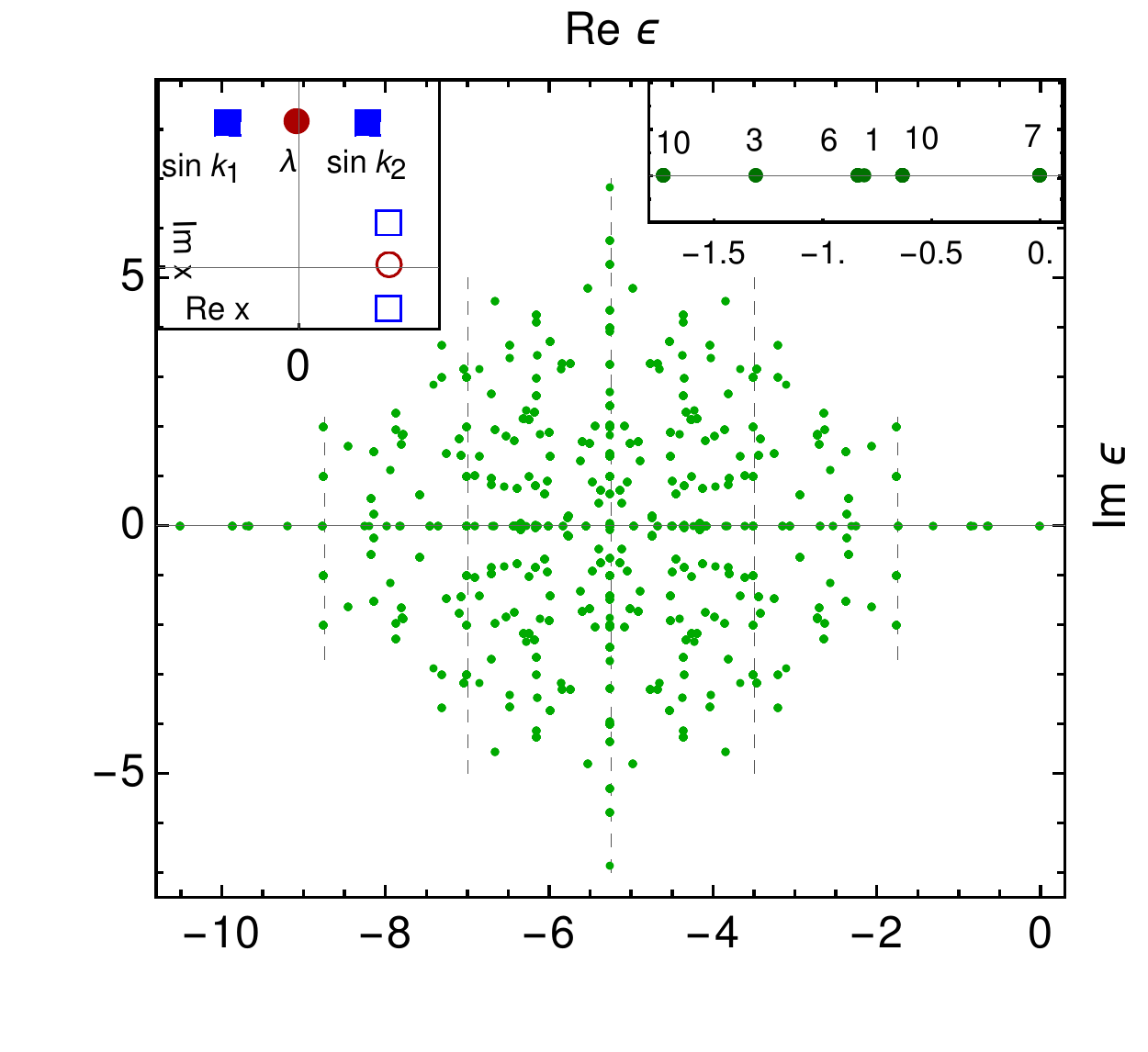}\\
\vspace{-6mm}
\caption{\label{pic}
Spectrum of the tight-binding model with dephasing for $L=6$,
$2\gamma=1.75$. Vertical dashed lines indicate multiples of
$-2\gamma$. As a result of the $\Pm\Tm$-symmetry\cite{PT} of the
Liouvillian the spectrum exhibits a $D_2$ point symmetry. 
Left inset: Schematic representation of the $1 k\Lambda$-string
solution of the Bethe equations, for the imaginary $u$-Hubbard model
(full symbols), and for the usual Hubbard chain with real $u$ (empty
symbols). Right inset: Close-up of the slowest decaying modes. The
numbers indicate the degeneracies, which are
consistent with translation from the lower to higher magnetization
sectors via the $\eta$-symmetry. For example, $10=5 \cdot 2$ is
obtained by counting the number of $\eta$-pairing descendant states in
sectors with higher numbers of particles (there are $L-1=5$), while
the factor of $2$ is due to a degeneracy in the sector $N=2$, $M=1$ of
$\eta$-pairing lowest weight states, which is a consequence of the
parity symmetry of the Hamiltonian (\ref{ImHubb}). The non-degenerate
state shown has $N=2M=6$ and is a singlet both with respect to parity
and the $\eta$-pairing.
}
\end{center}
\end{figure}

Spectral properties of the Hubbard Hamiltonian are commonly analyzed in
the framework of the so-called
string hypothesis\cite{Takahashibook,Takahashi,bookHub}, which  
assumes that, up to corrections that are exponentially small in system
size, the roots of all solutions form particular ``string'' patterns
in the complex plane. The structure of solutions of the imaginary-$u$
BAE is substantially different than in the usual Hubbard
model. Interestingly, there exists a class of string solutions
involving both $k$'s and $\Lambda$'s and is important for describing
the late time behavior. A single such ``$k$-$\Lambda$ string'' of length $m$
consists of $2m$ charge rapidities $k^{(m)}_{\alpha, j}$ and $m$ spin
rapidities $\Lambda_{\alpha,   j}^{(m)}$ such that for $\lambda^{(m)}_\alpha<0$
\begin{eqnarray}
k^{(m)}_{\alpha, j} &=& \arcsin(i \lambda^{(m)}_\alpha - (m - 2j +2)\gamma),
\nonumber\\
k^{(m)}_{\alpha,j+m} &=& \pi - \arcsin(i \lambda^{(m)}_\alpha + (m - 2j +2)\gamma) ,
\nonumber\\
\Lambda^{(m)}_{\alpha, j} &=& i\lambda^{(m)}_\alpha + \gamma (m+1 - 2j),~ 1\le j\le m. \label{T3}
\end{eqnarray}
For positive $\lambda^{(m)}_\alpha>0$ the structure of the string
solution is the same as (\ref{T3}) with the replacement
$\gamma\rightarrow-\gamma$. 
We stress that (\ref{T3}) are quite distinct from
$k$-$\Lambda$-strings in the usual Hubbard model~\cite{bookHub}:
the string centers are imaginary rather than real, and the $k$'s
enter as pairs ($k$, $\pi+k^*$) rather than as ($k$, $k^*$). An example
of a $k$-$\Lambda$ string solution in our model is shown
in the left inset of Fig.~\ref{pic}. In the usual Hubbard model a
$k$-$\Lambda$-string of length $m$ is a multi-particle bound state of
$m$ fermions with spin up and $m$ fermions with spin
down~\cite{bookHub,Essl1992}. In the present context $k$-$\Lambda$ 
string solutions to the BAE correspond to density matrices of our open
system, that have exponentially decaying off-diagonal matrix elements. 
For example, in the sector $N=2M=2$ the NESS-excitations $\Li\rho_m =
  \epsilon_m \rho_m,~\rho_m=\rho_m^\dg$ can be represented as $\rho_m = \sum_{x_1,x_2}
  \rho_m(x_1,x_2) \ket{0\ldots 1_{x_1}\ldots   0}\bra{0\ldots
    1_{x_2}\ldots 0}$, where
$\rho_m(x_1,x_2) = \begin{cases}
                    f(x_1,x_2),~x_1\ge x_2,\\
                    -f(x_1,x_2),~x_1<x_2
                   \end{cases}$ with
$$f(x_1,x_2)=i^{2(x_1-x_2)}(w+(-1)^{x_1+x_2}w^*)e^{-\xi_m |x_1-x_2|},$$
$w=e^{-i  q_m (x_1 + x_2)}$ and
$q_m = \pm \frac{\pi m}{L}$, $\xi_m = \arccosh \frac{\gamma}{\sin q_m}$, 
$1\leq m\leq\frac{L}{2}$.

For the particular subset of solutions of (\ref{BAE})
that consists only of $k$-$\Lambda$ strings we may use the
string hypothesis (\ref{T3}) to obtain the following set of
equations for the string centers $\lambda^{(n)}_\alpha$
\be
L f_m(\lambda^{(m)}_\alpha)=2\pi J^{(m)}_\alpha+\sum_{(m,\beta)}
\Theta_{nm}\Big(\frac{\lambda^{(m)}_\alpha-\lambda^{(n)}_\beta}{\gamma}\Big)\ .
\label{Takahashi}
\ee
Here $f_m(x)={\rm sgn}(x)\big(\pi-{\rm arcsin}(ix+m\gamma)+{\rm
  arcsin}(ix-m\gamma)\big)$, and $\Theta_{nm}(x)=2\theta(x/|n-m|+2)+\dots
+2\theta(x/n+m-2)+\theta(x/n+m)+(1-\delta_{n,m})\theta(x/|n-m|)$ with
$\theta(x)=2{\rm arctan}(x)$ is the same function as in the usual
  Hubbard model. The (half-odd) integers $J^{(m)}_\alpha$ have ranges
$|J^{(m)}_\alpha|\leq \frac{L-1}{2}-\frac{1}{2}
\sum_{n=1}\big(2{\rm min}(m,n)-\delta_{m,n}\big)M_n$.
Here $M_n$ is the number of $k$-$\Lambda$ strings of length $n$, and
$N=2M=\sum_{n=1}^\infty 2n M_n$.
The corresponding eigenvalues of the Liouvillian are real and given by
\bea
\epsilon&=&4\sum_{(m,\alpha)}{\rm Im}\sqrt{1-(i|\lambda^{(m)}_\alpha|-m\gamma)^2}
-2\gamma N.
\label{eigenvalues}
\eea
Moreover, studies of small systems $L\leq 8$ strongly suggest
that $k$-$\Lambda$ string solutions and their $\eta$-pairing
descendant states provide all slowly decaying NESS-excitations. In
Fig.~\ref{pic} we show the full spectrum of $\Li$ for $L=6$. The
states with ${\rm Re}(\epsilon)>-2\gamma$ are all given
by $k$-$\Lambda$ string solutions and their $\eta$-pairing
descendants, 
\emph{cf.} the right inset of Fig.~\ref{pic}. The situation for $L=8$ is analogous. Assuming this to hold in
general, we can obtain the eigenvalues of $\Li$ with the largest real
parts (i.e. the eigenvalues closest to zero) from the equations
(\ref{Takahashi}) for the string centers. For solutions consisting of
a single $k$-$\Lambda$ string $n$, i.e. $M_n=1$, $M_{j\neq n}=0$, we
find a sequence of eigenvalues with
\be
\epsilon_{n,j} =-\frac{1}{\gamma}\cdot\frac{2\pi^2
  (n+j-1)^2}{n L^2}+{\cal O}(L^{-4})\ ,\ 
j\in \mathbb{N}^+, \label{bands}
\ee
where we have assumed $n,|j|\ll L$. 
Let us denote the corresponding eigenstates by
$|n,j\rangle$. Using the $\eta$-pairing symmetry we can construct
degenerate states in the sector $N=2M=2k+2n$ of the form
\be
(\eta^\dagger)^k|n,j\rangle\ .
 \ee
This shows that {\it the spectrum of the dephasing model is gapless}
in the thermodynamic limit in any magnetization sector. 
For large but finite $L$ the smallest gap is given by $\epsilon_{1,1}$. 
The above construction carries over to general nonintegrable bipartite lattices in
the sense that NESS-excitations can be constructed from
two-particle states by acting with an appropriate power of
$\eta^\dagger$. In contrast to the usual Hubbard
model~\cite{unstableEtaPair}, perturbation theory in $1/\gamma$
suggests that $\eta$-pairing NESS-excitations are stable to typical
perturbations in the sense that they do not couple to states in the
complex part of the spectrum and retain real eigenvalues.

{\it Large $\gamma$ limit.--} 
It is known that in the limit of strong dephasing $\gamma\gg1$ the
late time dynamics of our system is described by a classical
stochastic process on the space of the diagonal density matrices, 
with an evolution operator that is equivalent to Hamiltonian of the
spin-1/2 Heisenberg chain~\cite{Cai}. This relation implies that the
NESS-excitations are gapless in any space dimension, as the lowest
lying excitation of the ferromagnet are gapless.
We have already commented that the $k$-$\Lambda$ solutions describing
the longest living NESS-excitations have exponentially decaying
off-diagonal matrix elements (they decay exponentially away from
diagonal with a scale determined from the solution of the BAE). 
In the large $\gamma$ limit further simplifications occur. In
particular the BAE (\ref{Takahashi}) reduce to Takahashi's equations
for the spin-1/2 Heisenberg ferromagnet:~\cite{Takahashibook} 
rescaling the rapidities
$\lambda^{(m)}_\alpha=\gamma \mu^{(m)}_\alpha$ and then taking
$\gamma\to\infty$ gives 
\bea
L \theta(\mu^{(m)}_\alpha/m)&=&2\pi J^{(m)}_\alpha+\sum_{(m,\beta)}
\Theta_{nm}\big({\mu^{(m)}_\alpha-\mu^{(n)}_\beta}\big)\ ,\nonumber\\
\epsilon&=&-\frac{1}{\gamma}\sum_{(m,\alpha)}\frac{2m}{m^2+(\Lambda^{(m)}_\alpha)^2}.
\label{TakahashiXXX}
\eea
The emergence of an effective description in terms of a Heisenberg
ferromagnet in the large-$\gamma$ regime of the dissipative model
should be contrasted to the large-$u$ expansion in the real-$u$
Hubbard model. Indeed, the low energy manifold for the Hubbard model
with large real interaction strength consists of configurations with
zero or one fermion per site only, while for the strongly dissipative
case the allowed configurations are those with zero or two fermions
per site (corresponding to the mostly diagonal density matrices). The
analogue of the large-$u$ expansion in the dissipative case gives
access to rapidly decaying modes with $\epsilon\approx-2\gamma N$.

{\it Relaxation dynamics.--} It has been shown~\cite{Esp2005,Eis2011}
that in the sector $N=2,~M=1$ the relaxation dynamics is diffusive,
both by means of spectral considerations~\cite{Esp2005} and by
studying the decay~\cite{Eis2011} $\ave{2a^\dagger_1 a_1 - 1} \propto
t^{-1/2}$ of an initially localized excitation $\ave{2a^\dagger_j a_j - 1}_{t=0} =
\delta_{j,1}$~\cite{DecayExplan}. By considering the off-diagonal
elements of the density matrix and using the duality between the
density matrices and correlation functions, one can easily calculate
the decay of coherences in the many-particle states of the
tight-binding model and obtain the dependence $\sim t^{-3/2}$ (similar
to a numerical result for the coherences in the $XXZ$ model with
dephasing obtained in Ref.~\onlinecite{Cai}). The long time relaxation
of many-particle states is influenced by the bound-state-like
NESS-excitations.

{\it XX model with dephasing.--} Most of our results apply also to the
spin-1/2 $XX$ chain with dephasing $H^{\mathrm{XX}}=\tfrac{1}{2}\sum_{j}^L
(\spl{j+1} \smi{j} + \spl{j} \smi{j+1})$,
$l^{\mathrm{XX}}_j=\sqrt{\gamma/2}\sz{j}$, where
$\spl{},\smi{},\sz{}$ are Pauli spin matrices. The $XX$ chain can be
mapped to a tight-binding model with dephasing by means of a
Jordan-Wigner transformation, but we now have to impose
periodic/antiperiodic (p/a) boundary conditions in the even/odd
magnetization sectors.\cite{bookNagaosa} Proceeding as before we eventually arrive at
an imaginary-$u$ Hubbard Hamiltonian (\ref{ImHubb}). However,
spin-$\sigma$ fermions now have periodic (antiperiodic) boundary
conditions, if their total number is even (odd). Altogether we
therefore have four distinct sectors (p,p), (p,a), (a,p), (a,a).
As a consequence of this the imaginary-$u$ Hubbard Hamiltonian does
not exhibit the full $SO(4)$ symmetry, but as $(\eta^+)^2$ acts
within a given sector, it commutes with the Hamiltonian. In spite of
the changed boundary conditions the model remains integrable. The
BAE are again of the form (\ref{BAE}), but we now have
$F_1=(-1)^{N-M-1}$, $F_2=(-1)^{N}$. Low-lying excitations can again be
analyzed by means of the string hypothesis (\ref{T3}) and the
equations (\ref{Takahashi}) for the string centers. 
The main difference to the dissipative tight-binding model is that in
the $XX$ case there is no closed form expression for the expectation values of
the $k$-point spin correlations functions, as the spins are non-local
in terms of Jordan-Wigner fermions.   

{\it Generalization to open boundaries.--} 
If in addition to dephasing there is influx/outflux of particles on the boundary sites, i.e. there are additional Lindblad operators
$l^{}_{1,2} = a_{1,L}^\dg,~l^{}_{3,4} = a_{1,L}$, or
$l^{\mathrm{XX}}_{1,2} = \spl{1,L},~l^{\mathrm{XX}}_{3,4} =
\smi{1,L}$, the Liouvillian in the thermofield language has the form
of the Hubbard model with imaginary interaction and with an imaginary
boundary magnetic field. The resulting model is again Bethe ansatz
solvable~\cite{Chineese}. We note that the current-carying NESS of such a
model has a simple explicit matrix product form \cite{Marko}. 

{\it Conclusions.--} We have shown that a dissipative tight-binding
model on a bipartite lattice can be mapped to a Hubbard model with
imaginary interaction strength. The NESS and the relaxational dynamics
at late times is related to the $\eta$-pairing symmetry of the Hubbard
model. In one spatial dimension we have used the Bethe ansatz solution
to derive exact results on the spectrum of the Liouvillian. Our result
pave the way to further studies of Bethe ansatz solvable quantum
dissipative systems.

This work was supported by the Slovenian Research Agency (ARRS)
under grants J1-5439 and N1-0025, the ERC grant OMNES (MVM and TP),
and by the EPSRC under grant EP/N01930X. FHLE and TP thank the Isaac
Newton Institute for Mathematical Sciences for hospitality and support
under grant EP/K032208/1. MVM is greatful to Marko Medenjak for
numerous discussions.


\begin{thebibliography}{99}
\bibitem{fermionicHS} Let us note that in general fermionic models in the double channel representation have a more complicated structure of the Hilbert space compared to the spin systems. The obtained simple form of the dissipator is due to simple form of the Lindblad operators: they conserve the particle number and are quadratic.
\end{thebibliography}
\end{document}